\documentstyle[aps,jpc,preprint]{revtex}
\tightenlines
\input epsf.sty

\begin{document}

\title{The Hydration Number of  Li$^+$ in 
Liquid Water\footnote{LA-UR-99-3360.}}

\author{Susan B. Rempe, Lawrence R. Pratt, Gerhard Hummer,  
Joel D. Kress, Richard L. Martin, and Antonio Redondo}

\address{Theoretical Division, Los Alamos National Laboratory, Los
Alamos, New Mexico 87545 USA}

\date{\today}

\maketitle

\begin{abstract}
A theoretical treatment based upon the quasi-chemical theory of
solutions predicts the most probable number of water neighbors in the
inner shell of a Li$^+$ ion in liquid water to be four.  The
instability of a six water molecule inner sphere complex relative to
four-coordinated structures is confirmed by an `ab initio' molecular
dynamics calculation.  A classical Monte Carlo simulation equilibrated
26 water molecules with a rigid six-coordinated Li(H$_2$O)$_6{}^+$
complex with periodic boundary conditions in aqueous solution.  With
that initial configuration for the molecular dynamics, the
six-coordinated structure relaxed into four-coordinated arrangements
within 112~fs and stabilized.  This conclusion differs from
prior interpretations of neutron and X-ray scattering results on
aqueous solutions.
\end{abstract}

\newpage

The hydration of ions in water is not only fundamental to physical
chemistry but also relevant to the current issue of selectivity of
biological ion channels.  In the context of potassium
channels\cite{kchannel:98,guidoni:99,laio:99}, for example, the free
energies for replacement of inner shell water ligands with peptide
carbonyls donated by proteins of the channel structure seem decisive
to the selectivity of the channel, specifically for preference of
K$^+$ over Na$^+$.  Studies to elucidate the thermodynamic features of
such inner shell exchange reactions require prior knowledge of the ion
hydration structures and energetics.

Unfortunately, our understanding of the inner hydration shell
structure of ions in water is not as clear as it might
be\cite{Friedman:85}.  The simplest and most favorable case to pursue
is the Li$^+$ solute.  Neutron scattering measurements on LiCl
solutions in liquid water have led to a firm conclusion that the
Li$^+$ ion has six near-neighbor water molecule
partners\cite{Friedman:85,Newsome:80,Enderby:81,Hunt:83,%
ichikawa:84,vanderMaarel:89,Howell:96}.  That result, however, has not
been entirely uniform across studies of similar aqueous
solutions\cite{cartailler:91,yamagami} containing Li$^+$ ions.  X-ray
scattering results have been interpreted similarly\cite{Radnai:81} to
indicate a hydration number of six, again with some
nonuniformity\cite{narten:73}.  In contrast, some spectroscopic
studies have suggested tetrahedral coordination of the Li$^+$ ion in
water \cite{Michaellian:78} and an array of physical chemical
inferences lend some support to that conclusion\cite{Ohtaki:93}.  On
the theoretical side, electronic structure calculations on the Li$^+$
ion with six water molecules predict a slightly, but distinctly, lower
energy for a structure with four inner shell and two outer shell water
molecules than for structures with six water molecules in the
innermost shell\cite{Feller:94,Feller:95}; results such as those seem
to be universally supported by other electronic structure
efforts\cite{bischof:97,tongraar:98}.  Simulations have produced a
range of results including both four and six inner shell water
neighbors with considerable statistical
dispersion\cite{Heinzinger:79,Mezei:81,Impey:83,Chandrasekhar:84,%
bounds:85,Zhu:91,Romero:91,Heinzinger:93,lee:94,toth:96,obst:96,%
Koneshan:98b}.  It is well recognized, of course, that simulations are
typically not designed to provide a sole determination of such
properties, though they do shed light on the issues determining the
hydration number of ions in water.

The theoretical scheme used here to address these problems for the
Li$^+$(aq) ion is based upon the quasi-chemical organization of
solution theory, which is naturally suited to these
problems\cite{pratt:98,feature,martin:97,pratt:99}.  The first step is
the study of the reactions

\begin{equation}
Li^+ + n H_2O \rightleftharpoons Li(H_2O)_n{}^+
\label{equilibria}
\end{equation}
\noindent
that combine $n$ water molecule ligands with the Li$^+$ ion in a
geometrically defined inner sphere under ideal gas conditions.  At a
subsequent step an approximate, physical description of the aqueous
environment surrounding these complexes is
included\cite{pratt:98,feature,martin:97,pratt:99}.  The geometric
definition of an inner sphere region enforces a physical balance in
this method.  The goal of this approach is to treat inner sphere
ligands explicitly, in molecular detail, but at the same time to
achieve a description of outer sphere hydration thermodynamics that is
consistent from one complex to another.  If minimum energy complex
geometries were to shift different numbers of ligands to outer sphere
regions, that would unbalance the thermodynamic description of the
hydration of the inner sphere materials.  For example, in the
quantitative implementation of the quasi-chemical approach we
specifically do not use the Li[(H$_2$O)$_4$][(H$_2$O)$_2$]$^+$ complex
cited above, with two water molecules outside the inner sphere, even
though this structure helpfully clarifies the physical issue.

Gas-phase thermochemical data required for the equilibria in
Eq.~(\ref{equilibria}) were obtained by electronic structure
calculations using the Gaussian98 programs with the B3LYP hybrid
density functional theory approximation\cite{G98}.  All structures
were fully optimized with a basis including polarization functions on
Li$^+$ (6-31G*) and both polarization and diffuse functions
(6-31++G**) on the oxygen and hydrogen centers.  At the optimum
geometry and with the same basis set, harmonic vibrational frequencies
of the clusters were calculated and atomic charges determined using
the ChelpG capability in Gaussian98.  Partition functions were then
calculated, thus providing a determination of the free energy changes
of the equilibria in Eq.~(\ref{equilibria}) due to atomic motions
internal to the clusters within the harmonic approximation.

Interactions of these complexes with the external aqueous
environment\cite{feature} were treated with a dielectric model
following the previous study of the hydrolysis of the ferric
ion\cite{martin:97}.  Classic electrostatic interactions based upon
the ChelpG partial atomic charges were the only solution-complex
interactions treated; in particular, repulsive force (overlap)
interactions were neglected based on the expectation that they
make a secondary contribution to the thermodynamic properties
considered here.  The external boundary of the volume enclosed by
spheres centered on all atoms defined the solute molecular surface.
The sphere radii were those determined empirically by Stefanovich and
Truong\cite{Stefanovich}, except R$_{Li^+}$=2.0~\AA\ for the lithium
ion.  Because the lithium ion is well buried by the inner shell
waters, slight variations of the lithium radius were found to be
unimportant.  The value R$_{Li^+}$=2.0~\AA\ was identified as slightly
larger than the nearest Li-O distances and significantly smaller than
the Li-O distances (3.5 -- 4.0~\AA) for second shell pairs.

Results of the calculations are summarized in Fig.~1.  Geometry
optimization of each of the $n$-coordinated clusters confirms that the
inner shell structures used in these calculations are not necessarily
the lowest energy structures for a given number of water neighbors.
Although a tetrahedral cluster of inner shell water molecules is the
lowest energy structure for Li(H$_2$O)$_4{}^+$, a cluster with five
inner shell water molecules is slightly higher in energy than a
cluster with one outer shell and four inner shell water molecules.
Similarly, the lowest energy cluster with six water molecules contains
four inner shell water molecules arranged tetrahedrally and two outer
shell water molecules.

Fig.~1 shows that the $n$=4 inner sphere cluster has the lowest free
energy for a dilute (p=1~atm) ideal gas phase.  Adjustment of the
concentration of water molecules to the value $\rho_W$ = 1~g/cm$^3$,
to match the normal density of liquid water, changes the most favored
cluster to the one with $n$=6 inner shell water molecules.  Outer
sphere interactions described by the dielectric model progressively
destabilize the larger clusters, as they should since larger numbers
of water molecules are being treated explicitly as members of the
inner shell.  As a consequence of including the outer sphere
contributions, the final position of minimum free energy is returned
to the $n$=4 structure, with the $n$=3 complex predicted to be next
most populous in liquid water at T=298.15~K and p=1~atm.  The mean
hydration number predicted by this calculation is ${\overline n}$=4.0.

The current quasi-chemical prediction for the absolute hydration free
energy of the Li$^+$ ion under these conditions is -128~kcal/mol, not
including any repulsive force (packing) contributions.  An extreme
increase of R$_{Li^+}$ to 2.65~\AA\ raises this value to about
-126~kcal/mol, showing that the theoretical results are insensitive to
the ion radius, as remarked above.  Experimental values are
-113~kcal/mol\cite{marcus:94}, -118~kcal/mol\cite{conway}, and
-125~kcal/mol\cite{Friedman:73}, converted to this standard state.
This dispersion of experimental values for the absolute hydration free
energy of the Li$^+$ (aq) ion is accurately mirrored in the dispersion
of reference values adopted for the absolute hydration free energy of
the H$^+$ (aq) ion.  Inclusion of repulsive force contributions would
reduce the present calculated value slightly.  Furthermore, Li$^+$(aq)
is believed to have a strongly structured second hydration
shell\cite{Impey:83}, which is treated only approximately in this
calculation.  Nevertheless, this level of agreement between
calculation and experiment is satisfactory.

We additionally emphasize that the Li(H$_2$O)$_n{}^+$ complexes are
treated in the harmonic approximation, although fully quantum
mechanically.  The low-$n$ clusters might have more entropy than is
being accounted for by the harmonic approximation.  If this were the
case, then low-$n$ clusters would be more populous than currently
represented.  This would likely raise the theoretical value also.

To further test the $n$=4 prediction, `ab initio' molecular dynamics
calculations were carried out utilizing the VASP program\cite{vasp1}.
Two checks established the consistency for these problems between the
electronic structure calculations described above and the energetics
involved in the molecular dynamics calculations.  First, the electron
density functional alternative implemented in VASP\cite{pw91} was
checked by comparing the electronic structure results obtained with
the B3LYP hybrid electron density functional and the PW91 generalized
gradient approximation exchange-correlation functional, using the
Gaussian98 program and the same basis sets.  As expected,
satisfactory agreement was observed in the binding energies for
sequential addition of a water molecule to the Li(H$_2$O)$_n{}^+$
clusters.  Then the issues of pseudo-potentials and basis set were
checked by optimizing cluster geometries with the VASP program and
comparing to the results obtained for the same problems with
Gaussian98.  Again agreement was observed.  For example, both
procedures predicted the same lowest energy six-coordinated structure,
the characteristic Li[(H$_2$O)$_4$][(H$_2$O)$_2$]$^+$ cluster, with
nearly identical geometries.

To initiate the `ab initio' molecular dynamics calculation, the
optimum $n$=6 inner sphere structure, rigidly constrained, was first
equilibrated with 26 water molecules under conventional Monte Carlo
liquid simulation conditions for liquid water, including periodic
boundary conditions.  This system of one Li$^+$ ion and 32 water
molecules was then used as an initial configuration for the 
molecular dynamics calculation.  As shown in Fig.~2, the initial $n$=6
structure relaxed to stable $n$=4 alternatives within 112~fs.  The
results of longer molecular dynamics calculations will be reported
later.

The `ab initio' molecular dynamics and the quasi-chemical theory of
liquids exploit different approximations and produce the same
conclusion here.  This {\em agreement} supports the prediction that
Li$^+$(aq) has four inner shell water ligands at infinite dilution in
liquid water under normal conditions.  This prediction differs from
interpretations of neutron and X-ray scattering data on aqueous
solutions.

The conditions studied by these calculations and those targeted in the
neutron scattering work do not match perfectly, particularly with
regard to Li$^+$ concentration.  Nevertheless, the theoretical methods
are straightforward and physical, and, moreover, the distinct methods
used here conform in their prediction of hydration number.  Therefore,
it will be of great importance for future work to fully resolve the
differences between calculations and scattering experiments for these
problems.

This work was supported by the US Department of Energy under contract
W-7405-ENG-36 and the LDRD program at Los Alamos.

\begin{figure} [b!]
\epsfxsize=5.2in 
\epsfysize=5.2in 
\epsfbox{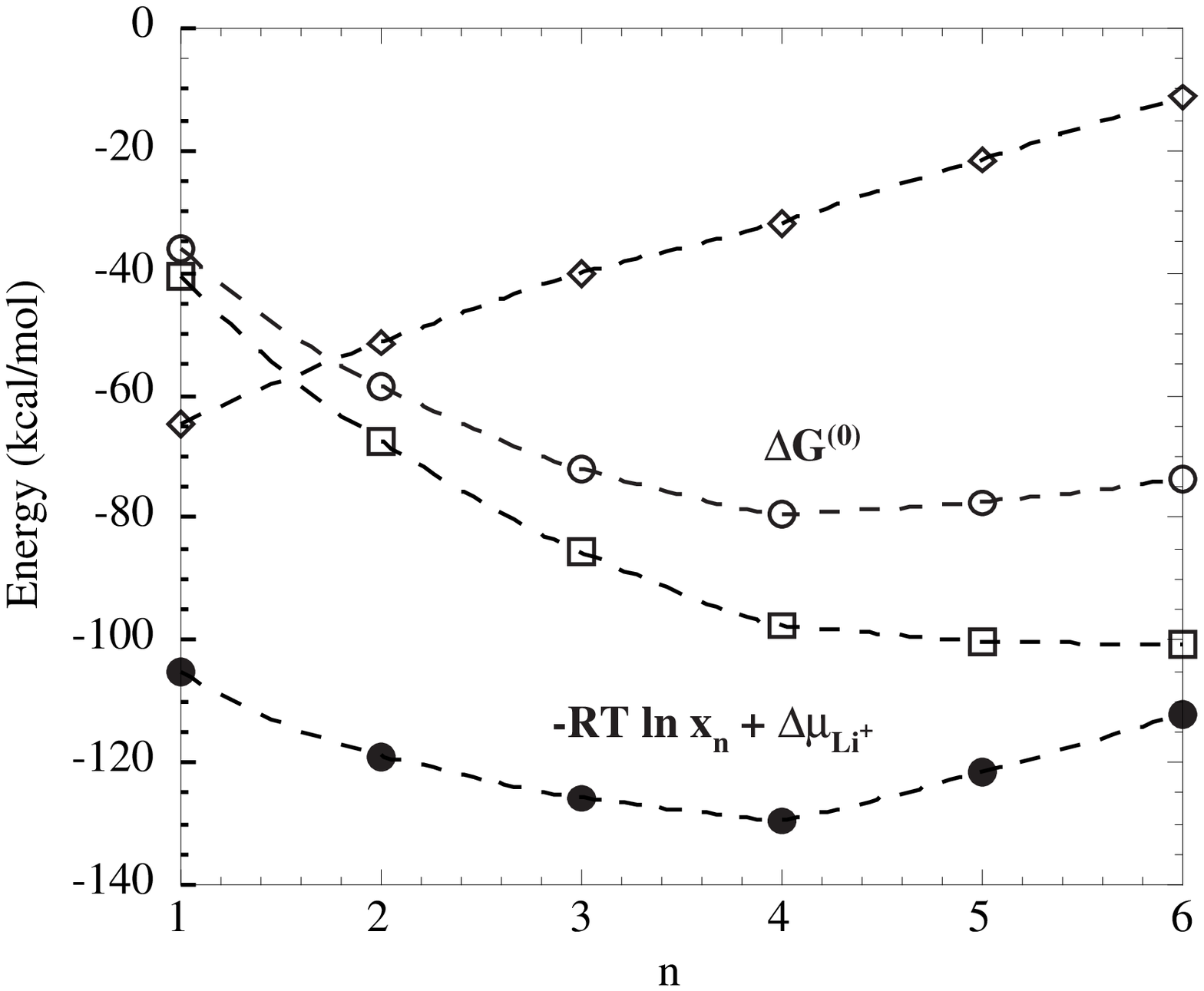}
%\vspace{0.2in}
\caption{Free energies for Li$^+$ ion hydration  in
liquid water as a function of the number of inner shell water
neighbors at T=298.15~K.  The results marked $\Delta$G$^{(0)}$ (open
circles) are the free energies predicted for the reaction Li$^+$ +
$n$H$_2$O = Li(H$_2$O)$_n{}^+$ under standard ideal
conditions, including p = 1~atm.  The minimum value is at $n$=4.  The
next lower curve (squares) incorporates the replacement free energy
-nRT $\ln$(RT$\rho_W$/1~atm) that adjusts the concentration of water
molecules to the normal concentration of liquid water, $\rho_W$ =
1~g/cm$^3$ so that RT$\rho_W$ = 1354~atm\protect{\cite{martin:97}}.
The minimum value is at $n$=6.  The topmost graph (diamonds) plots
$\mu^*_{Li(H_2O)_n{}^+}-n\mu^*_{H_2O}$, the external-cluster
contributions obtained from the standard dielectric
model\protect{\cite{feature,martin:97}}.  The bottommost results
(solid circles) are the final, net values.  The label provides the
quasi-chemical expression of these net values
\protect{\cite{pratt:98,pratt:99}} with $x_n$ the  fraction of lithium 
ions having $n$ inner shell water neighbors and $\Delta\mu_{Li^+}$ the
interaction part of the chemical potential of the lithium ions.  This
graph indicates that the $n$=4 inner sphere structure is most probable
in liquid water under normal conditions.}
\end{figure}

\begin{figure}
\hspace{1.0in}
\epsfbox{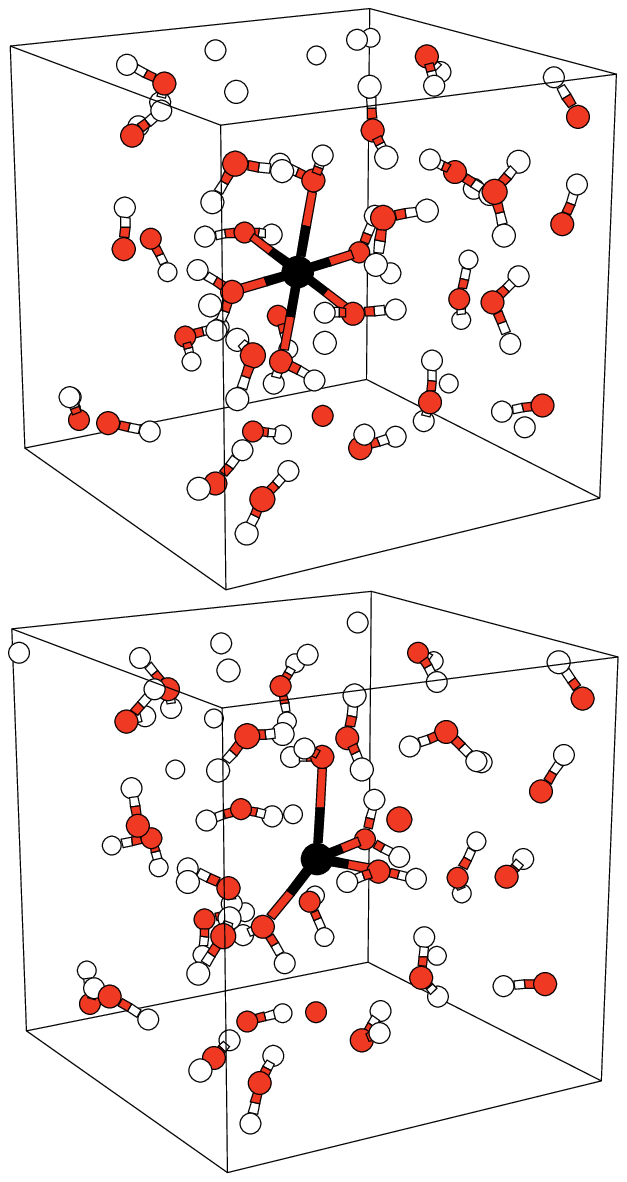}
\caption{Structures from molecular dynamics
calculations based upon a gradient-corrected electron density
functional description of the interatomic forces.  The ions were
represented by ultrasoft pseudopotentials\protect{\cite{vasp2}} and a
kinetic energy cutoff of 396~eV, which was found satisfactory in
related calculations\protect{\cite{sprik}}, limited the plane wave
expansions . The top panel is the configuration used as an initial
condition.  A hexa-coordinate inner sphere structure, rigidly
constrained, was equilibrated with 26 additional water molecules by
Monte Carlo calculations using classical model force fields and
assuming a partial molar volume of zero.  The bottom panel is the
structure produced 112~fs later.  The bonds identify water oxygen
atoms within 2.65~\AA\ of the Li$^+$ ion.  The hydrogen, lithium, and
oxygen atoms are shown as open, black, and gray circles,
respectively.}
\end{figure}


\begin{thebibliography}{10}

\bibitem{kchannel:98}
 Doyle, D.~A.; Cabral, J.~M.; Pfuetzner, R.~A.; Kuo, A.~L.; Gulbis,
J.~M.;  Cohen, S.~L.;  Chait, B.~T.;  MacKinnon, R.
\newblock {\em Science} {\bf 1998}, 280, 69--77.

\bibitem{guidoni:99}
 Guidoni, L.; Torre, V.; Carloni, P.
\newblock {\em Biochem.} {\bf 1999}, 38, 8599--8604.

\bibitem{laio:99}
 Laio, A.; Torre, V.
\newblock{\em Biophys. J.} {\bf 1999}, 76, 129--148.

\bibitem{Friedman:85}
Friedman, H.~L. 
\newblock {\em Chem. Scr.} {\bf 1985}, 25, 42--48.

\bibitem{Newsome:80}
Newsome, J.~R.; Neilson, G.~W.; Enderby, J.~E.
\newblock {\em J. Phys. C: Solid St. Phys.} {\bf 1980}, 13, L923--L926.

\bibitem{Enderby:81}
 Enderby, J.~E.;  Neilson, G.~W.
\newblock {\em Rep. Prog. Phys.} {\bf 1981}, 44, 593--653.

\bibitem{Hunt:83}
Hunt, J.~P.; Friedman, H.~L. 
\newblock {\em Prog. Inorg. Chem.} {\bf 1983}, 30, 359--387.

\bibitem{ichikawa:84}
 Ichikawa, K.; Kameda, Y.; Matsumoto, T.; Masawa, M.
\newblock {\em J. Phys. C: Solid State Phys.} {\bf 1984},
 17, L725--L729.

\bibitem{vanderMaarel:89}
 van der Maarel, J.~R.~C.; Powell, D.~H.; Jawahier, A.~K.;
 Leyte-Zuiderweg, L.~H.; Neilson, G.~W.; Bellissent-Funel,
 M.~C.
\newblock {\em J. Chem. Phys.} {\bf 1989}, 90, 6709--6715.

\bibitem{Howell:96}
 Howell, I.; Neilson, G.~W.
\newblock {\em J. Phys.: Condens. Matter} {\bf 1996}, 8, 4455--4463.

\bibitem{cartailler:91}
 Cartailler, T.; Kunz, W.; Turq, P.; Bellissent-Funel, M.~C.
\newblock {\em J. Phys.: Condens. Matter} {\bf 1991}, 3, 9511--9520.

\bibitem{yamagami}
 Yamagami, M.; Yamaguichi, T.; Wakita, H.; Misawa, M.;
\newblock {\em J. Chem. Phys.} {\bf 1994}, 100, 3122--3126.
 
\bibitem{Radnai:81}
Radnai, T.; P{\'a}link{\'a}s, G.; Szasz, G.~I.; Heinzinger, K.
\newblock {\em Z. Naturforsch. A} {\bf 1981}, 36, 1076--1082.

\bibitem{narten:73}
 Narten, A.~H.; Vaslow, F.; Levy, H.~A.
\newblock {\em J. Chem. Phys.} {\bf 1973}, 85, 5017--5023.

\bibitem{Michaellian:78}
Michaellian, K.~H.; Moskovits,  M.
\newblock {\em Nature} {\bf 1978}, 273, 135 -- 136.

\bibitem{Ohtaki:93}
Ohtaki, H.; Radnai, T.;
\newblock {\em Chem. Rev.} {\bf 1993}, 93, 1157--1204.

\bibitem{Feller:94}
Feller, D.; Glendening, E.~D.; Kendall, R.~A.; Peterson, K.~A.;
\newblock {\em J. Chem. Phys.} {\bf 1994}, 100, 4981--4997.

\bibitem{Feller:95}
Feller, D.; Glendening, E.~D.; Woon, D.~E.; Feyereisen, M.~W.;
\newblock {\em J. Chem. Phys.} {\bf 1995}, 103, 3526--3542.

\bibitem{bischof:97}
Bishof, G.; Silbernagel, A.; Hermansson, K.; Probst, M.
\newblock {\em Int. J. Quant. Chem.} {\bf 1997}, 65, 803--816.

\bibitem{tongraar:98}
Tongraar, A.; Liedl, K.~R.; Rode, B.~M.
\newblock {\em Chem. Phys. Letts.} {\bf 1998}, 286, 56--64.

\bibitem{Heinzinger:79}
Heinzinger, K.; P{\'a}link{\'a}s,  G.
\newblock {\em The Chemical Physics of Solvation}
\newblock Elsevier, Amsterdam, 1985; pp. 313.

\bibitem{Mezei:81}
Mezei, M.;  Beveridge, D.~L.
\newblock {\em J. Chem. Phys.} {\bf 1981}, 74, 6902--6910.

\bibitem{Impey:83}
 Impey, R.~W.; Madden, P.~A.; McDonald, I.~R.
\newblock {\em J. Phys. Chem.} {\bf 1983}, 87, 5071--5083.

\bibitem{Chandrasekhar:84}
Chandrasekhar, J.; Spellmeyer, D.~C.;  Jorgensen, W.~L. 
\newblock {\em J. Am. Chem. Soc.} {\bf 1984}, 106, 903--910.

\bibitem{bounds:85}
 Bounds, D.~G.
\newblock {\em Mol. Phys.} {\bf 1985}, 54, 1335--1355.

\bibitem{Zhu:91}
 Zhu, S.~B.; Robinson, G.~W.
\newblock {\em Z. Naturforsch. A} {\bf 1991}, 46, 221--228.

\bibitem{Romero:91}
Romero, C.
\newblock {\em J. Chim. Phys.} {\bf 1991}, 88, 765--777.

\bibitem{Heinzinger:93}
Heinzinger, K.
\newblock in {\em Water-Biomolecule Interactions}
\newblock eds.  M. U. Palma, M. B. Palma-Vittorelli, and F. Patak:
SIF, Bologna, 1993.  p. 23.

\bibitem{lee:94}
 Lee, S.~H.; Rasaiah, J.~C.
\newblock {\em J. Chem. Phys.} {\bf 1994}, 101, 6964--6974.

\bibitem{toth:96}
Toth, G.
\newblock {\em J. Chem. Phys.} {\bf 1996}, 105, 5518--5524.

\bibitem{obst:96}
Obst, S; Bradaczek, H.
\newblock {\rm J. Phys. Chem.} {\bf 1996}, 100, 15677--15687.

\bibitem{Koneshan:98b}
Koneshan, S.; Rasaiah, J.~C.; Lynden-Bell, R.~M.; Lee, S.~H.
\newblock {\em J. Phys. Chem. B} {\bf 1998}, 102, 4193--4204.

\bibitem{pratt:98}
 Pratt, L.~R.; LaViolette, R.~A.
\newblock {\em Mol. Phys.} {\bf 1998}, 94, 909.

\bibitem{feature}
Hummer, G.; Pratt, L.~R.; Garc{\'{\i}}a, A.~E.
\newblock {\em J. Phys. Chem. A} {\bf 1998}, 102, 7885--7895.

\bibitem{martin:97}
Martin, R.~L.; Hay, P.~J.; Pratt, L.~R.
\newblock {\em J. Phys. Chem. A} {\bf 1998}, 102, 3565--3573.

\bibitem{pratt:99}
Pratt, L.~R.; Rempe, S.~B.
\newblock {\em Quasi-Chemical Theory and Implicit Solvent Models for
Simulations} {\bf 1999}, LA-UR-99-3125.

\bibitem{G98}
M.~J. Frisch, G.~W. Trucks, H.~B. Schlegel, G.~E. Scuseria, M.~A.
Robb, J.~R.  Cheeseman, V.~G. Zakrzewski, J.~A. Montgomery, R.~E.
Stratmann, J.~C. Burant, S.~Dapprich, J.~M. Millam, A.~D. Daniels,
K.~N. Kudin, M.~C. Strain, O.~Farkas, J.~Tomasi, V.~Barone, M.~Cossi,
R.~Cammi, B.~Mennucci, C.~Pomelli, C.~Adamo, S.~Clifford,
J.~Ochterski, G.~A. Petersson, P.~Y. Ayala, Q.~Cui, K.~Morokuma, D.~K.
Malick, A.~D. Rabuck, K.~Raghavachari, J.~B. Foresman, J.~Cioslowski,
J.~V. Ortiz, B.~B. Stefanov, G.~Liu, A.~Liashenko, P.~Piskorz,
I.~Komaromi, R.~Gomperts, R.~L. Martin, D.~J. Fox, T.~Keith, M.~A.
Al-Laham, C.~Y. Peng, A.~Nanayakkara, C.~Gonzalez, M.~Challacombe,
P.~M.~W. Gill, B.~G.  Johnson, W.~Chen, M.~W. Wong, M.~Head-Gordon,
E.~S. Replogle, and J.~A.  Pople.
\newblock {\em Gaussian 98 (Revision A.2)}.
\newblock Gaussian, Inc., Pittsburgh PA, 1998.

\bibitem{Stefanovich}
 Stefanovich, E.~V.; Truong, T.~N.
\newblock {\em Chem. Phys. Lett.} {\bf 1995}, 244, 65--74.

\bibitem{marcus:94}
Marcus, Y.
\newblock {\em Biophys. Chem.} {\bf 1994}, 51, 111--127.

\bibitem{conway}
Conway, B.~E.
\newblock {\em J. Soln. Chem.} {\bf 1978}, 7, 721--770.
\bibitem{Friedman:73}
 Friedman, H.~L., and Krishnan, C.~V., in {\em Water A Comprehensive
Treatise} Vol.~3, edited by F. Franks: Plenum Press, New York, 1973, 
p. 1.

\bibitem{vasp1}
Kresse, G.; Hafner, J.
\newblock {\em Phys. Rev. B} {\bf 1993}, 41, 558.

\bibitem{pw91} Perdew, J.; Burke, K.; Wang, Y. 
\newblock {\em Phys. Rev. B} {\bf 1996}, 54, 16533.

\bibitem{vasp2}
Kresse, G.; Hafner, J. 
\newblock {\em J. Phys.: Condens. Mat.} {\bf 1994}, 6, 8245.

\bibitem{sprik}
Sprik, M.; Hutter, J.; Parrinello, M. 
\newblock {\em J. Chem. Phys.} {\bf 1996}, 105, 1142.
\end{thebibliography}
\end{document}